\documentclass[conference]{IEEEtran}
\IEEEoverridecommandlockouts

\usepackage{cite}
\usepackage{amsmath,amssymb,amsfonts}
\usepackage{algorithmic}
\usepackage{graphicx}
\usepackage{textcomp}
\usepackage{subcaption}  
\usepackage{dblfloatfix}

\usepackage{xcolor}
\usepackage{float}
\usepackage{comment}
\usepackage{booktabs} 
\usepackage{array}    
\usepackage{afterpage}

\makeatletter
\newcommand{\linebreakand}{%
  \end{@IEEEauthorhalign}
  \hfill\mbox{}\par
  \mbox{}\hfill\begin{@IEEEauthorhalign}
}
\makeatother

    \author{\IEEEauthorblockN{1\textsuperscript{st} Arek Berç Gökdağ}
    \IEEEauthorblockA{\textit{DEIB} \\
    \textit{Politecnico di Milano}\\
    Milan, Italy \\
    arekberc.gokdag@polimi.it}
    \and
    \IEEEauthorblockN{2\textsuperscript{nd} Given Name Surname}
    \IEEEauthorblockA{\textit{dept. name of organization (of Aff.)} \\
    \textit{name of organization (of Aff.)}\\
    City, Country \\
    email address}
    \and
    \IEEEauthorblockN{3\textsuperscript{rd} Given Name Surname}
    \IEEEauthorblockA{\textit{dept. name of organization (of Aff.)} \\
    \textit{name of organization (of Aff.)}\\
    City, Country \\
    email address}
    \linebreakand 
    \IEEEauthorblockN{4\textsuperscript{th} Given Name Surname}
    \IEEEauthorblockA{\textit{dept. name of organization (of Aff.)} \\
    \textit{name of organization (of Aff.)}\\
    City, Country \\
    email address}
    \and
    \IEEEauthorblockN{5\textsuperscript{th} Given Name Surname}
    \IEEEauthorblockA{\textit{dept. name of organization (of Aff.)} \\
    \textit{name of organization (of Aff.)}\\
    City, Country \\
    email address}
    \and
    \IEEEauthorblockN{6\textsuperscript{th} Given Name Surname}
    \IEEEauthorblockA{\textit{dept. name of organization (of Aff.)} \\
    \textit{name of organization (of Aff.)}\\
    City, Country \\
    email address}
    }

\author{
\IEEEauthorblockN{Arek Berç Gökdağ}
\IEEEauthorblockA{\textit{DEIB} \\
\textit{Politecnico di Milano}\\
Milan, Italy \\
arekberc.gokdag@polimi.it}
\vspace{3ex}
\IEEEauthorblockN{Michele Zhu}
\IEEEauthorblockA{\textit{DEIB} \\
\textit{Politecnico di Milano}\\
Milan, Italy \\
michele.zhu@polimi.it}
\and

\IEEEauthorblockN{Silvia Mura}
\IEEEauthorblockA{\textit{DEIB} \\
\textit{Politecnico di Milano}\\
Milan, Italy \\
silvia.mura@polimi.it}
\vspace{3ex}
\IEEEauthorblockN{Maurizio Magarini}
\IEEEauthorblockA{\textit{DEIB} \\
\textit{Politecnico di Milano}\\
Milan, Italy \\
maurizio.magarini@polimi.it}

\and

\IEEEauthorblockN{Antonio Coviello}
\IEEEauthorblockA{\textit{DEIB} \\
\textit{Politecnico di Milano}\\
Milan, Italy \\
antonio.coviello@polimi.it}
\vspace{3ex}
\IEEEauthorblockN{Umberto Spagnolini}
\IEEEauthorblockA{\textit{DEIB} \\
\textit{Politecnico di Milano}\\
Milan, Italy \\
umberto.spagnolini@polimi.it}
}

\begin{document}

\title{Low-Complexity CNN-Based Classification of Electroneurographic Signals\\
{}}

\maketitle

\begin{abstract}
Peripheral nerve interfaces (PNIs) facilitate neural recording and stimulation for treating nerve injuries, but real-time classification of electroneurographic (ENG) signals remains challenging due to constraints on complexity and latency, particularly in implantable devices.
This study introduces MobilESCAPE-Net, a lightweight architecture that reduces computational cost while maintaining and slightly improving classification performance. Compared to the state-of-the-art ESCAPE-Net, MobilESCAPE-Net achieves comparable accuracy and F1-score with significantly lower complexity, reducing trainable parameters by 99.9\% and floating point operations per second by 92.47\%, enabling faster inference and real-time processing. Its efficiency makes it well-suited for low-complexity ENG signal classification in resource-constrained environments such as implantable devices.
\end{abstract}

\begin{IEEEkeywords}
ENG signals, Classification, Peripheral nerve interfaces.
\end{IEEEkeywords}

\section{Introduction}

The peripheral nervous system (PNS) transmits sensory and motor information between the central nervous system and the body. Damage to the PNS can cause peripheral neuropathy (PN), leading to sensory, motor, and autonomic impairments with long-term functional and socio-economic consequences. Mechanical injuries are a major cause of PN, contributing to disabilities that affect individuals and healthcare systems worldwide~\cite{[4]caillaud2019peripheral}.

Although the PNS can regenerate, nerve injury recovery is often incomplete, resulting in lasting impairments. Traditional treatments have limited effectiveness, increasing interest in bioelectronic medicine, particularly implantable peripheral nerve interfaces (PNIs)~\cite{paperasp}. These devices record electroneurographic (ENG) signals and enable neural decoding and stimulation (ND\&S), showing promising therapeutic potential~\cite{[5]Denison65}. 

A major challenge in ND\&S systems is the real-time classification of ENG signals, as the complexity of nerve signal propagation makes it difficult to capture meaningful activity with low latency. Classification is essential for interpreting neural intent or physiological events and delivering appropriate stimulation. Minimizing processing delays is particularly critical for restoring sensory and motor functions, where responses must occur within the $300\,$ms threshold, beyond which patients begin to perceive a delay~\cite{[10][11]Controller}. Overcoming these challenges is key to improve neural interfacing and advancing bioelectronic medicine.

Various classification methods have been explored for electroencephalography (EEG) and electromyographic (EMG) signals, which share similarities with ENG signals. These approaches include statistical models~\cite{martinek2021advanced}, machine learning (ML) methods~\cite{hosseini2020review}, and deep learning (DL) frameworks~\cite{10530475}. However, statistical models and ML methods often struggle with the significant noise and distortion in ENG, EEG, and EMG signals, leading to lower classification performance compared to DL methods~\cite{hosseini2020review}, which reduces their feasibility for use in ND\&S systems.

Deep learning techniques, particularly convolutional neural networks (CNNs), have been applied to ENG signal classification, including recordings from the sciatic nerve of Long-Evans rats~\cite{10530475,kohSelectivePeripheralNerve2020journalArticle}. The study in~\cite{10530475} compared several models, including ENGNet (adapted from EEGNet~\cite{lawhern2018eegnet}), Long Short-Term Memory (LSTM), and inception time models, with ENGNet achieving high classification performance while maintaining a simple design. Similarly,~\cite{kohSelectivePeripheralNerve2020journalArticle} introduced ESCAPE-Net, where raw ENG signals are preprocessed into spatiotemporal signatures (detailed in Sec. \ref{sect:Preproc}) for classification.

While DL techniques generally offer superior classification accuracy compared to statistical and traditional ML methods~\cite{sanei2021eeg}, their high computational demands~\cite{10530475, kohSelectivePeripheralNerve2020journalArticle} pose challenges for real-time applications. This limitation underscores the need for more efficient models capable of reducing computational overhead while maintaining classification performance.

To address this issue, this work introduces MobilESCAPE-Net, a novel architecture designed to reduce computational cost while preserving robust classification performance significantly. MobilESCAPE-Net is validated against ESCAPE-Net, which serves as a baseline using the same dataset from~\cite{SP3/JRZDDR_2023}. 

Numerical results demonstrate that a key advantage of MobilESCAPE-Net is its substantial reduction in model size and computational complexity, achieving a $99.92\,\%$ decrease in parameters, requiring only $68\,$k, and a $92.47\,\%$ reduction in floating point operations per second (FLOPs), requiring just $81.8\,$MFLOPs. Despite this significant decrease in model size and computational overhead, MobilESCAPE-Net not only matches but slightly outperforms ESCAPE-Net in both accuracy and F1-score. This efficiency makes MobilESCAPE-Net particularly well-suited for low-latency, resource-constrained environments, demonstrating the effectiveness of its optimized architecture.

The rest of the paper is structured as follows. Section~\ref{sect:systemModel} describes the system model under consideration, while Sec.~\ref{sect:Preproc} details the preprocessing of the ENG signal. Section~\ref{sect:Class} presents the classification methods, including both state-of-the-art and the proposed approach. Numerical results are provided in Sec.~\ref{sect:results}, and conclusions are drawn in Sec.~\ref{sect:conclusion}.

\subsubsection*{Notation} Lower-case letters $\mathbf{x}$ refer to vectors and the notation $x_{l,k}$ refers to the $(l,k)$th element. The symbols $||\cdot||$, $\mathbb{E}$ and $\mathbb{R}$ refer to the Euclidean norm operator, expectation operator, and to the real number set, respectively. 

\section{System Model}\label{sect:systemModel}

\begin{figure}[!t]
    \centering
    \includegraphics[width=0.95\linewidth]{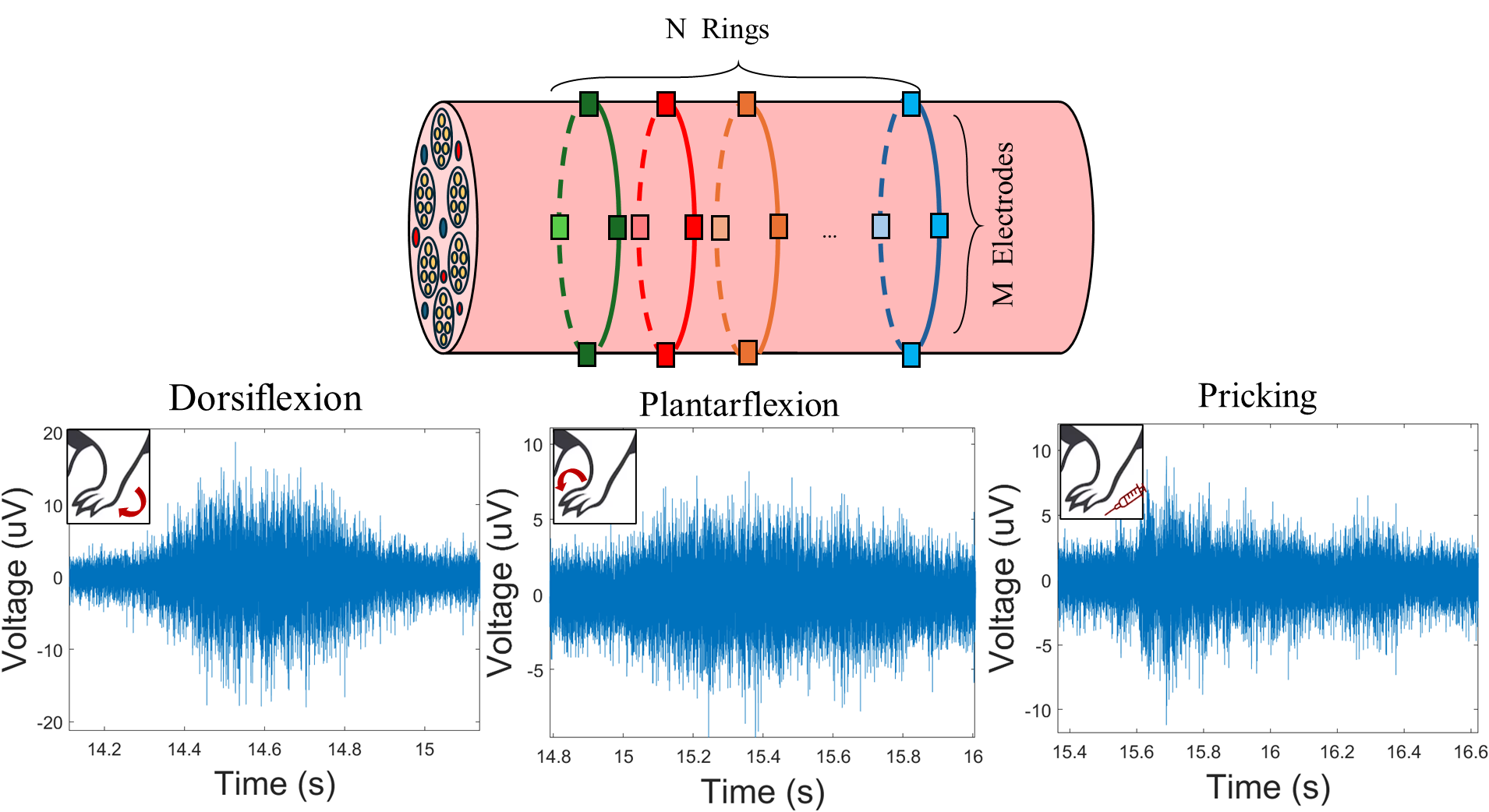}
    \caption{Cuff electrode model with example ENG signals recorded for dorsiflexion, plantarflexion, and pricking.}
    \label{fig:systemModel}
\end{figure}

Nerves consist of tightly packed axon bundles that facilitate the transmission of electrical ENG signals between the central nervous system and peripheral tissues, playing a crucial role in maintaining both signal conduction and structural stability. These ENG signals travel in an afferent direction to convey sensory input to the central nervous system or in an efferent direction to deliver motor commands to muscles~\cite{[4]caillaud2019peripheral}. To capture the aggregated neural activity generated by individual axons, we consider the cuff electrode illustrated in Fig.~\ref{fig:systemModel}, which is modeled as a cylindrical structure composed of
$N$ rings, each containing $M$ electrodes to enhance the redundancy of the measured data. Consequently, the total number of electrodes in the cuff electrode is given by $L$$\,=\,$$ N$$\,\times\,$$ M$. Assuming the presence of 
$K$ uncorrelated ENG sources, represented by 
$\mathbf{s}(t) \in \mathbb{R}^{K \times 1}$, 
the signal ${y}_{l}$ received by the $l$th electrode can be modeled as~\cite{10530475}
\begin{align} 
{y}_{l}(t) = \sum_{k=1}^K{h}_{l,k}\,\,{s}_k(t-\tau_{l,k}) + {v}_l(t) + {u}(t), \end{align} 
where ${u}(t)$ represents the i.i.d. additive white Gaussian noise common to all electrodes and ${v}_{l}(t)$ denotes the interference signal. This interference arises from the cumulative effect of EMG signals generated by surrounding muscles during contractions and common to all the electrodes, along with electrode-specific artifacts~\cite{10530475}. Therefore, the interference signals exhibit a nonzero correlation, $\mathbb{E}[{v}_{l+1}(t)\,\,{v}_l(t)] \neq 0, \forall l$. 
The source signal $\mathbf{s}(t)$ combines efferent and afferent neural activity and is modeled as the sum of multiple compound action potentials (CAP). CAPs are extracellular spikes observed in nerve cuff recordings, resulting from the synchronous activation of small groups of nerve fibers~\cite{kohSelectivePeripheralNerve2020journalArticle}. The parameter $\tau_{l,k}$ denotes the propagation delay associated with the 
$k$th source as perceived by the $l$th electrode. The lead field $h_{l,k}$ is defined as in~\cite{10530475}
\begin{align} h_{l,k} = -\frac{1}{4 \pi \sigma ||\mathbf{q}_{l} - \mathbf{q}_k||^2} \end{align} 
with $\sigma$ representing the conductivity and $\mathbf{q}_{l}$ and $\mathbf{q}_k$ representing the position of the $l$th electrode and the position of the $k$th source , respectively, with respect to an arbitrary reference system. 

\section{ENG Signal Preprocessing}\label{sect:Preproc}

This section describes the preprocessing steps designed to enhance the quality of noisy ENG signals, which require a robust processing pipeline. A key aspect of this process is the detection of CAPs, which appear as distinct spikes in the ENG signal and are essential for assessing neural activity and extracting meaningful signal features.

The dataset used in this work~\cite{SP3/JRZDDR_2023} includes recordings from nine Long-Evans rats (Rat2–Rat10) subjected to dorsiflexion, plantarflexion, and pricking stimuli. Each rat was subjected to an approximately 180-second experiment per class,  with each experiment containing around 100 stimulus periods. Stimuli were applied manually in sync with a 70 BPM metronome to maintain consistent timing. 

\begin{figure}[!t]
    \centering
    \begin{minipage}{0.48\columnwidth}
        \centering
        \includegraphics[width=0.9\linewidth]{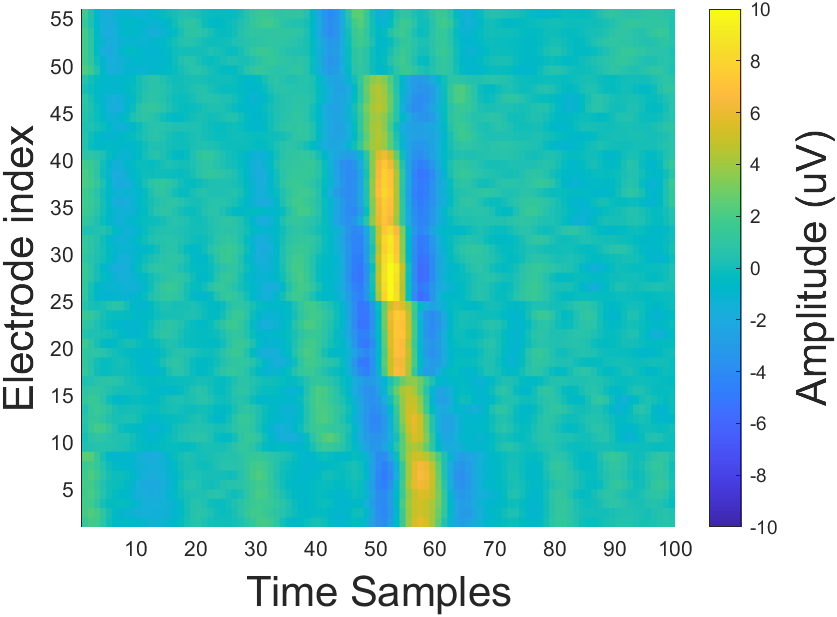}
        \subcaption{}
        \label{fig:subplot1_s}
    \end{minipage} \hfill
    \begin{minipage}{0.48\columnwidth}
        \centering
        \includegraphics[width=0.9\linewidth]{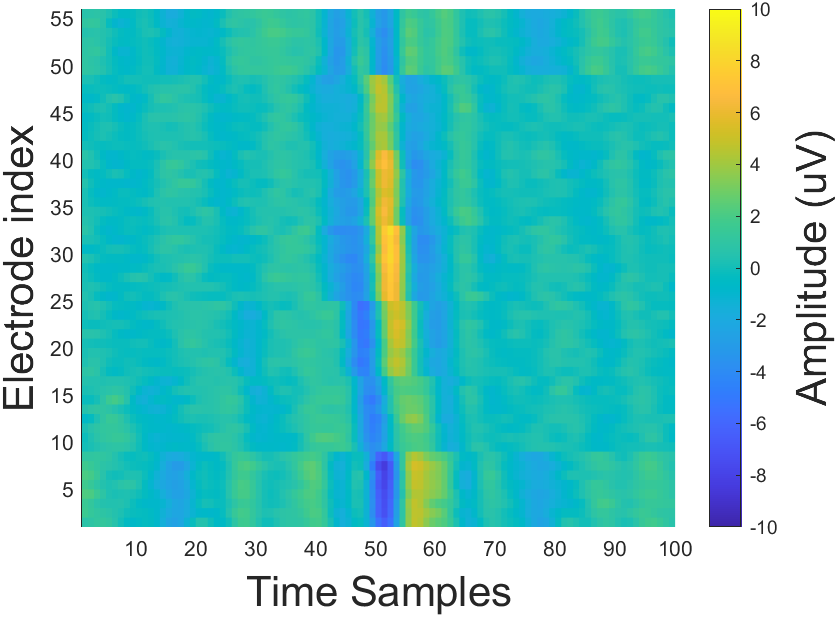}
        \subcaption{}
        \label{fig:subplot2_s}
    \end{minipage}

    \vspace{0.15cm}

    \begin{minipage}{0.48\columnwidth}  
        \centering
        \includegraphics[width=0.9\linewidth]{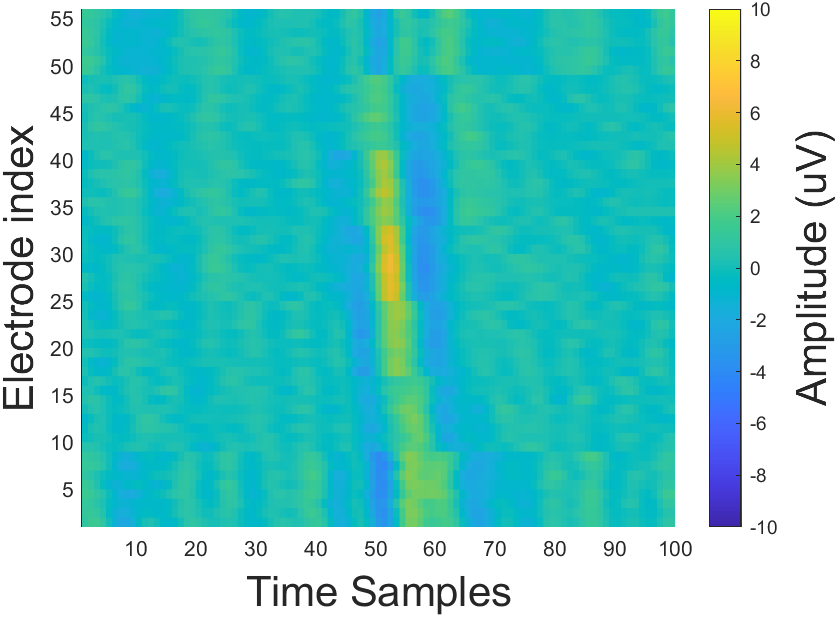}
        \subcaption{}
        \label{fig:subplot3_s}
    \end{minipage}
    \caption{Spatiotemporal signatures for dorsiflexion (a), plantarflexion (b), and pricking (c). Each image represents average of 40 spatiotemporal signatures, illustrating the distribution of neural activity across electrodes over time.}
    \label{fig:spatiotemporal_images}
\end{figure}

The proposed framework builds upon the methodology presented in~\cite{kohSelectivePeripheralNerve2020journalArticle}, introducing modifications to the processing sequence, bandpass filter parameters, and CAP detection criteria. The process consists of the following key stages:

\begin{itemize}
\item \textit{Removal of extreme values.} The signal ${y}_l(t)$, containing all the time samples, is clipped within the range of $\pm$40\,$\mu$V to eliminate extreme outliers.
\item \textit{Bandpass Filtering.} A $6$th-order Butterworth filter is applied to retain the frequency components of the signal between $800\,$Hz and $5\,$kHz. This filtering step helps reducing the unwanted distortion ${v}_l(t)$, mainly related to EMG activity that usually occurs below $500\,$Hz, with harmonics reaching up to $800\,$Hz. A $5\,$kHz cutoff helps minimizing the electrode artifacts at higher frequencies~\cite{kohSelectivePeripheralNerve2020journalArticle}.
\item \textit{Tripolar Referencing.} It estimates common-mode noise ${{u}}(t)$ from two reference signals ${y}_l(t)$ and ${y}_{l'}(t)$ and filters it from the target signal to isolate neural activity~\cite{struijkTripolarNerveCuff1995conferencePaper}. In our setup, it is applied across the cuff array using reference electrodes from the outermost rings.
\begin{figure}[!t]
    \centering
    \includegraphics[width=1\linewidth]{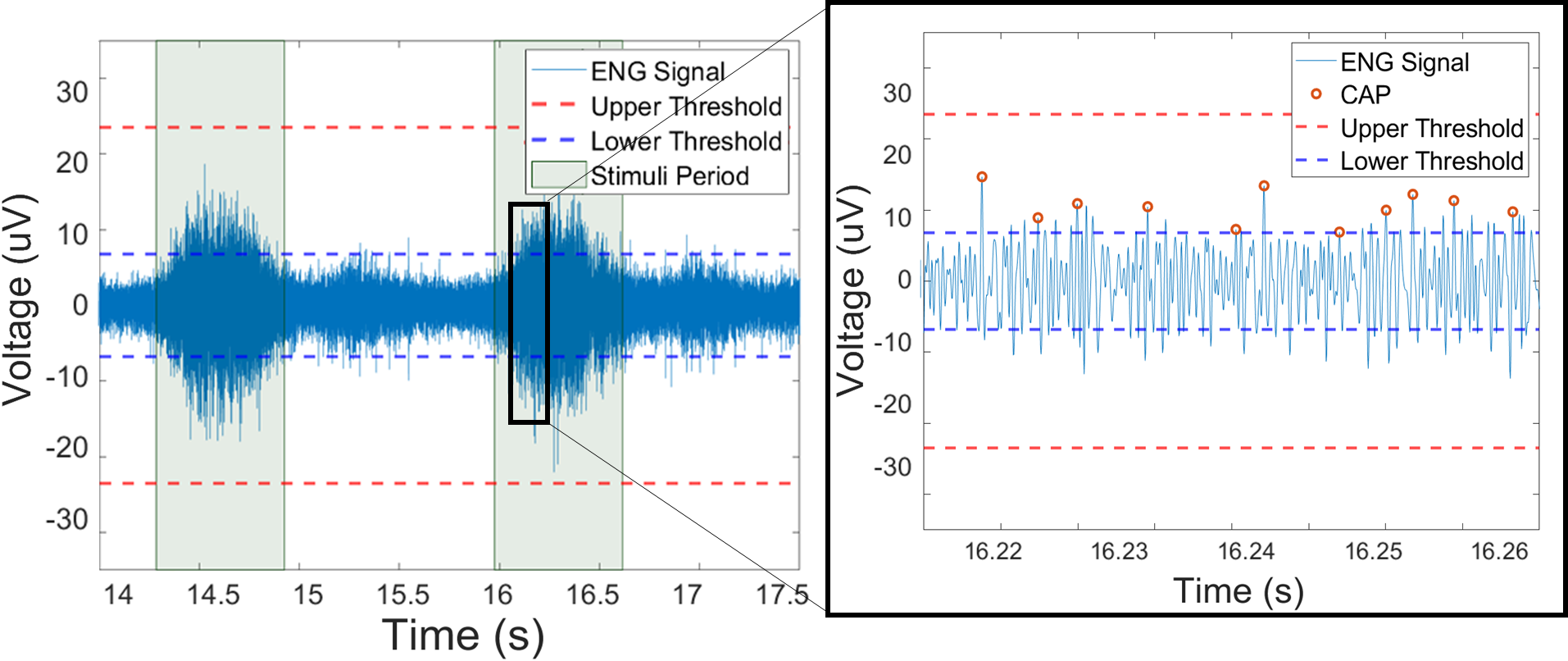}
    \caption{ENG signal recorded during dorsiflexion in Rat 6. The left part displays the estimated activation intervals, while the right part provides a zoomed-in view with the detected CAPs.}
    \label{fig:capest}
\end{figure}

\item \textit{Activity Interval Estimation.} The ENG signal consists of periods where stimulation is applied and not applied. To differentiate them, the moving average of absolute signal values is calculated, with peaks indicating the centers of stimulation periods. Data from the middle ring channels serves as a reference. Figure~\ref{fig:capest} shows the partitioning of the ENG signal, comprising neural recordings during stimulation The signal is segmented into activity and inactivity periods, effectively isolating the most energetically significant segments corresponding to neural responses.

\item \textit{Signal-to-noise ratio (SNR).} By calculating the power during all the stimulus applied (stimulus-on) periods, collected in $\mathcal{T}_{\text{on}}$ and all the not applied (stimulus-off) periods $(\mathcal{T}_{\text{off}})$ for each $l$th channel, the average SNR is then obtained by averaging the ratio over all \(L\) channels as
\[
\label{eq:snr}
\mathrm{SNR} = \frac{1}{L} \sum_{l=1}^{L} \frac{\| y_l (t) \|^2_{\mathcal{T}_{\text{on}}}}{\| y_l (t) \|^2_{\mathcal{T}_{\text{off}}}}.
\]

\begin{table}[!b]
    \centering
    \caption{SNR in dB for different rats and classes in the considered dataset.}
    \resizebox{0.7\columnwidth}{!}{
    \begin{tabular}{lccc}
        \toprule
        \textbf{Rat} & \textbf{Dorsiflexion} & \textbf{Plantarflexion} & \textbf{Pricking} \\
        \midrule
        Rat4 & 1.81 & 2.44 & 2.42 \\
        Rat5 & 2.03 & 2.12 & 2.83 \\
        Rat6 & 3.59 & 1.24 & 1.59 \\
        Rat7 & 0.23 & 1.97 & 0.87 \\
        Rat8 & 0.93 & 1.96 & 0.55 \\
        Rat9 & 1.6 & 0.41 & 0.61 \\
        \bottomrule
    \end{tabular}
    }
    \label{tab:snr_matrix}
\end{table}

This calculation is performed for each rat and class, considering all stimulation-applied and non-applied periods. The number of stimulation-applied periods varies based on the specific animal and stimulus used, resulting in an average stimulation window of 0.64 seconds.
Table~\ref{tab:snr_matrix} presents the average SNR for the retained rats. Datasets from subjects with particularly low SNR values (Rat2, Rat3, and Rat10) are excluded to ensure that the analyses are based on datasets with clearly defined stimulus periods.


\item \textit{CAP detection.} CAPs are identified by detecting signal peaks within thresholds derived from the signal's standard deviation and median values according to~\cite{kohSelectivePeripheralNerve2020journalArticle}.  To improve specificity, a temporal exclusion criterion with $3\,$ms window is applied, eliminating closely spaced peaks and reducing ripple artifacts.

\item \textit{Spatio-Temporal Signature Definition.} For each CAP, the temporal location from the middle ring data centers the signature. Signals from all $L$ electrodes are organized into a matrix, where each column represents simultaneous recordings at a given time step. These matrices are combined to form a unified spatiotemporal signature capturing both spatial and temporal features of the neural response. Figure~\ref{fig:spatiotemporal_images} illustrates the signatures for three distinct ENG stimuli: dorsiflexion, plantarflexion, and pricking, highlighting their differences to facilitate classification. The resulting images are then used as input for ESCAPE-Net and MobilESCAPE-Net to perform classification.

\end{itemize}

\section{ENG signals Classification }\label{sect:Class}

This section presents MobilESCAPE-Net, an optimized alternative to ESCAPE-Net, maintaining classification performance with lower computational cost. After preprocessing (Sec.~\ref{sect:Preproc}), spatiotemporal signatures are used to classify stimuli into dorsiflexion, plantarflexion, and pricking.
\begin{figure*}[!b]
    \centering
    \includegraphics[width=0.88\linewidth]{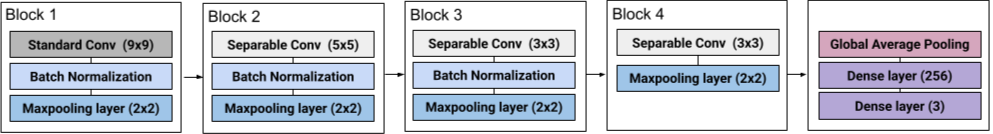}
    \caption{MobilESCAPE-Net architecture.}
    \label{fig:mobilescapenet}
\end{figure*}

\subsection{ESCAPE-Net}
ESCAPE-Net is a CNN consisting of three convolutional layers, two max pooling layers for feature extraction, and a fully connected layer for classification~\cite{kohSelectivePeripheralNerve2020journalArticle}. As detailed in Table \ref{tab:escape_mobile_arch}, its architecture features progressively smaller convolutional kernels (8×8, 4×4, and 2×2), each with 64 filters and rectified linear unit (ReLU) activation functions, followed by a fully connected layer with 256 nodes and a three-class softmax output. While its high parameter count supports strong performance, it also incurs substantial computational and memory overhead, limiting real-time applicability. A key contributor to the high parameter count is the max pooling configuration, where padding and stride settings preserve feature map dimensions, constraining dimensionality reduction. Additionally, the flattening layer converts feature maps into a high-dimensional vector, further increasing parameters in both the extraction and classification layers.


\subsection{MobilESCAPE-Net}\label{sect:Mobil}
MobilESCAPE-NET is introduced to enhance efficiency and decrease complexity by targeting four key aspects of the network's architecture.

First, the max pooling configuration is adjusted by increasing the stride from $1$ to $2$ pixels per step, ensuring alignment with the $2$$\,\times\,$$2$ kernel size to maintain information integrity. Larger strides or kernel sizes are avoided to prevent excessive reduction in feature map dimensions. Additionally, the padding scheme are changed from \textit{same} to \textit{valid} padding, allowing for controlled down-sampling while preserving essential spatial information.
Second, we simplify the padding scheme of the convolutional layers by replacing even-sized kernels ($8$$\,\times\,$$8$, $4$$\,\times\,$$4$ and $2$$\,\times\,$$2$) with odd-sized ones ($9$$\,\times\,$$9$, $5$$\,\times\,$$5$ and $3$$\,\times\,$$3$). This modification ensures symmetric padding, which enhances the consistency and reliability of feature extraction throughout the network. 
Third, to mitigate the issue of model size while maintaining performance, we implement parameter optimization through the use of depthwise and pointwise convolutions~\cite{howardMobileNetsEfficientConvolutional2017preprinta}. These techniques reduce the number of trainable parameters, resulting in a more efficient network while preserving high classification accuracy. Finally, we assess the integration of Global Average Pooling (GAP)~\cite{linNetworkNetwork2014preprint} as an alternative to the flattening layer of the ESCAPE-net, achieving a significant reduction in the parameter count. Moreover, batch normalization is integrated to stabilize the training process, accelerate convergence, and improve overall efficiency by regulating network activations.

\begin{figure*}[!t]
    \centering
    \begin{minipage}{0.32\textwidth}
        \centering
        \includegraphics[width=0.85\textwidth]{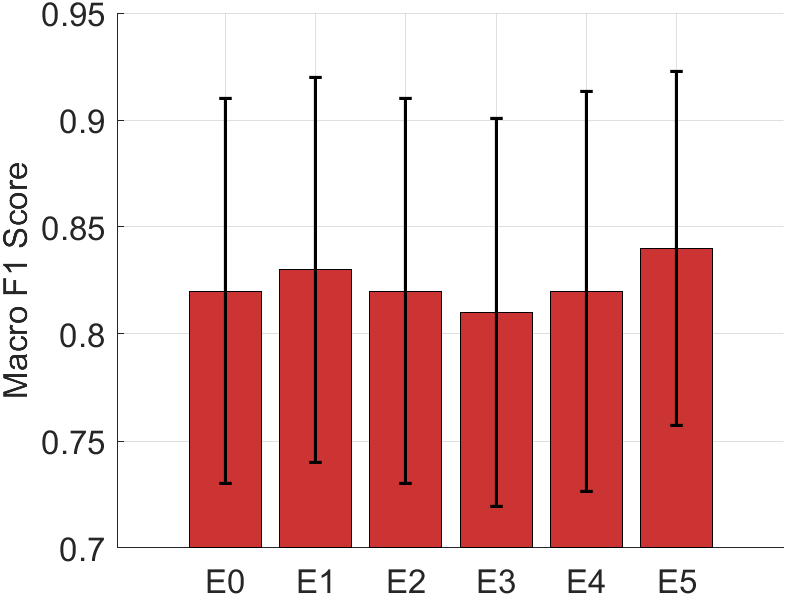}  
        \subcaption{}  
        \label{fig:subplot1_ablation}
    \end{minipage}
    \begin{minipage}{0.32\textwidth}
        \centering
        \includegraphics[width=0.85\textwidth]{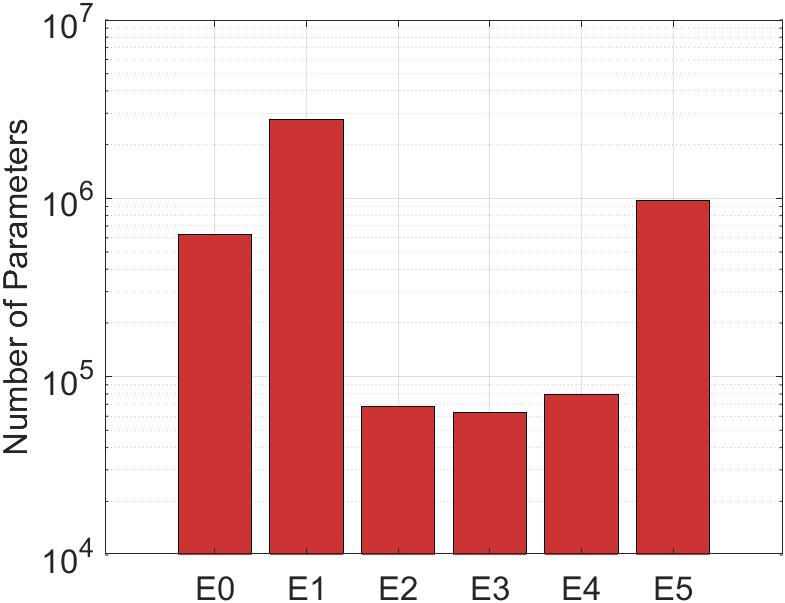}  
        \subcaption{}  
        \label{fig:subplot2_ablation}
    \end{minipage}
    \begin{minipage}{0.32\textwidth}
        \centering
        \includegraphics[width=0.85\textwidth]{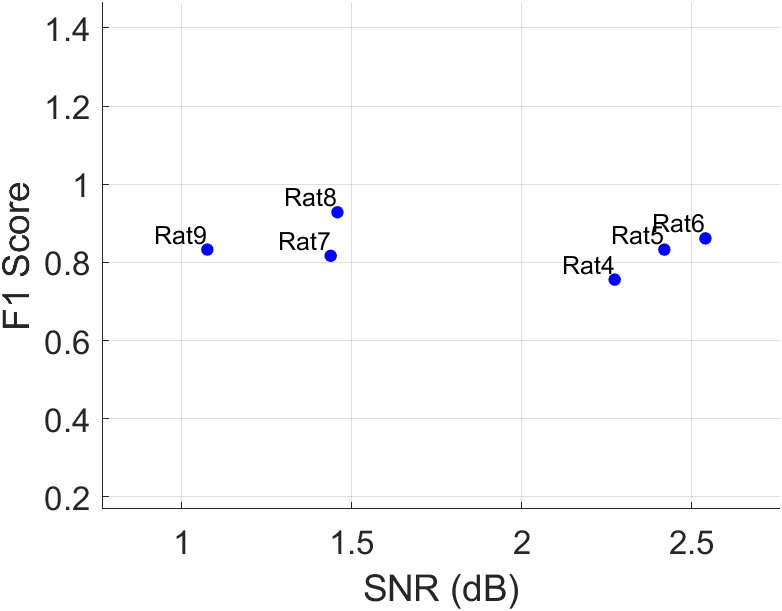}  
        \subcaption{}  
        \label{fig:subplot3_model_optimization_study}
    \end{minipage}
    \caption{Results of the model optimization study:  (a) Validation of the Macro F1-Score (\%) across the studied configurations in Table~\ref{tab:model_optimization_results}, with error bars indicating variance. (b) The number of parameters for each configuration, highlighting model complexity differences. (c) Scatter plot of SNR in dB vs. MobileEscapeNet F1-Score for different rats.}
    \label{fig:model_optimization_study}
\end{figure*}

The final MobilESCAPE-Net model, presented in Fig.~\ref{fig:mobilescapenet} and Table~\ref{tab:escape_mobile_arch}, exhibits a highly compact architecture with remarkably fewer trainable parameters, reinforcing its efficiency and practicality.

\begin{table}[ht]
\centering
\caption{Comparison of ESCAPE-Net and MobileESCAPE-Net architectures.}
\resizebox{\columnwidth}{!}{%
\begin{tabular}{@{}lll@{}}
\toprule
\textbf{Layer}                                & \textbf{ESCAPE-Net}                           & 
\textbf{MobileESCAPE-Net} \\
\toprule
Conv                              & Std. 8$\times$8, 64 filters, ReLU        & Std.  9$\times$9, 64 filters, ReLU \\
Batch Normalization                      & -               & Yes\\
MaxPool                       & 2$\times$2, stride 1                & 2$\times$2, stride 2 \\
\midrule
Conv                             & Std.  4$\times$4, 64 filters, ReLU        & Depthwise 5$\times$5 \\ 
                                              &                                   & Pointwise 1$\times$1, 64 filters, ReLU \\

Batch Normalization                      & -               & Yes\\
MaxPool                       & 2$\times$2, stride 1                & 2$\times$2, stride 2 \\
\midrule
Conv                             & Std.  2$\times$2, 64 filters, ReLU        & Depthwise 3$\times$3 \\ 
                                              &                                   & Pointwise 1$\times$1, 64 filters, ReLU  \\
Batch Normalization                      & -               & Yes\\
MaxPool                       & -                         & 2$\times$2, stride 2 \\
\midrule
Conv                             & -                          & Depthwise 3$\times$3 \\ 
                                              &                                   & Pointwise 1$\times$1, 128 filters , ReLU  \\
MaxPool                       & -                          & 2$\times$2, stride 2 \\
\midrule
Flattening                           & Yes                               & -- \\
Global Average Pooling               & --                                & Yes \\
Fully Connected                      & 256 nodes, ReLU                   & 256 nodes, ReLU \\
Output                               & 3 nodes, Softmax                  & 3 nodes, Softmax \\
\bottomrule
\end{tabular}
}
\label{tab:escape_mobile_arch}
\end{table}

\section{Numerical Results}\label{sect:results}
This section presents numerical results for ENG stimulus classification using ESCAPE-Net and the proposed MobilESCAPE-Net. First, the hyperparameter optimization process for MobilESCAPE-Net, along with performance evaluation, is outlined. Then, a comparative analysis of ESCAPE-Net and MobilESCAPE-Net is conducted, assessing classification accuracy, F1-score, macro F1-score, and model complexity. In this context, accuracy reflects correct classifications, while the F1-score is the harmonic mean of precision and recall. The macro F1-score provides an equal-weighted average of F1-score across all classes. Meanwhile, model complexity is evaluated based on trainable parameters and FLOPs.





\subsection{Optimization of MobilESCAPE-NET hyperparameters}\label{sect:ablation}
A k-fold cross-validation approach is employed to ensure a rigorous evaluation of model performance. Initially, $15\%$ of the dataset is set aside as a test set, reserved exclusively for final model assessment. The remaining $85\%$ is partitioned into five folds, with four utilized for training and one for validation in each iteration. This methodology mitigates bias associated with a single train-validation split and enhances the reliability of generalization estimates. To improve model robustness, pixel intensities are normalized prior to training, and the Adam optimizer with early stopping is implemented to mitigate overfitting. Furthermore, an empirical study on three subjects dataset with high-SNR  (Rat4, Rat5 and Rat6) is conducted to systematically assess architectural modifications.

\begin{table}[t!]
    \centering
    \caption{Model Optimization Study Most Significant Results.}
    \resizebox{\columnwidth}{!}{
    \begin{tabular}{l l c c}
        \toprule
        \textbf{Exp. ID} & \textbf{Modification} & \textbf{Kernel Sizes} & \textbf{\# of Filters} \\
        \midrule
        $E_0$  &  & 9×9, 5×5, 3×3, 3×3  & 64, 64, 64, 128 \\
        $E_1$  & $E_0$ + Remove Block 3  & 9×9, 5×5, 3×3  & 64, 64, 128\\
        $E_2$  & $E_0$ + GAP instead of flattening & 9×9, 5×5, 3×3, 3×3  & 64, 64, 64, 128\\
        $E_3$  & $E_1$ + GAP instead of flattening & 9×9, 5×5, 3×3  & 64, 64, 128 \\
        $E_4$  & $E_2$ + Larger Kernels  & 15×15, 7×7, 5×5, 3×3  & 64, 64, 128  \\
        $E_5$  & $E_2$ + More Filters  & 9×9, 5×5, 3×3, 3×3  & 256, 512, 512, 512  \\
        \bottomrule
    \end{tabular}
    }
    \label{tab:model_optimization_results}
\end{table}
While various architectural modifications were examined to optimize MobileESCAPE-Net, only the most significant results are reported here. As detailed in Table~\ref{tab:model_optimization_results} and Fig.~\ref{fig:model_optimization_study}, the model optimization process begins with the base model $E_0$, corresponding to the architecture shown in Fig. \ref{fig:mobilescapenet}, but using a flattening layer instead of GAP. Removing the third block (resulting in $E_1$) increases the Macro F1 score by 1.2\%, reaching 
0.83, but at the cost of increased complexity, requiring 2.7M parameters. Substituting the flattening layer with GAP in the model $E_2$ significantly reduces the parameter count by approximately 89\% while maintaining a similar Macro F1 score. Further removal of the third block in $E_3$ leads to a slight performance drop to 0.81 Macro F1 score. Adjusting the kernel size ($E_4$) results in a minimal F1 score improvement but slightly increases complexity. Finally, increasing the number of convolutional filters in $E_5$ achieves a Macro F1 score of 0.84, but with a 35\% increase in model complexity compared to $E_0$. In conclusion, the $E_2$ configuration of MobilESCAPE-Net is selected as the optimal balance between minimal complexity and classification performance, achieving a Macro F1 score of 0.82 with only 68k parameters, maintaining performance despite a 
89\% reduction in model size.

Figure~\ref{fig:subplot3_model_optimization_study} presents a scatter plot depicting the correlation between SNR (dB) and the F1-score achieved by MobileESCAPE-Net. The results show that the MobilESCAPE-NET ($E_2$) performs robustly across various SNR levels, with an average F1-score of $0.84$ and a standard deviation of $0.1$, demonstrating its capability to handle variations in signal quality effectively.

 \begin{figure}[!b]
    \centering
    \begin{minipage}{0.65\columnwidth}
        \centering
        \includegraphics[width=0.8\columnwidth]{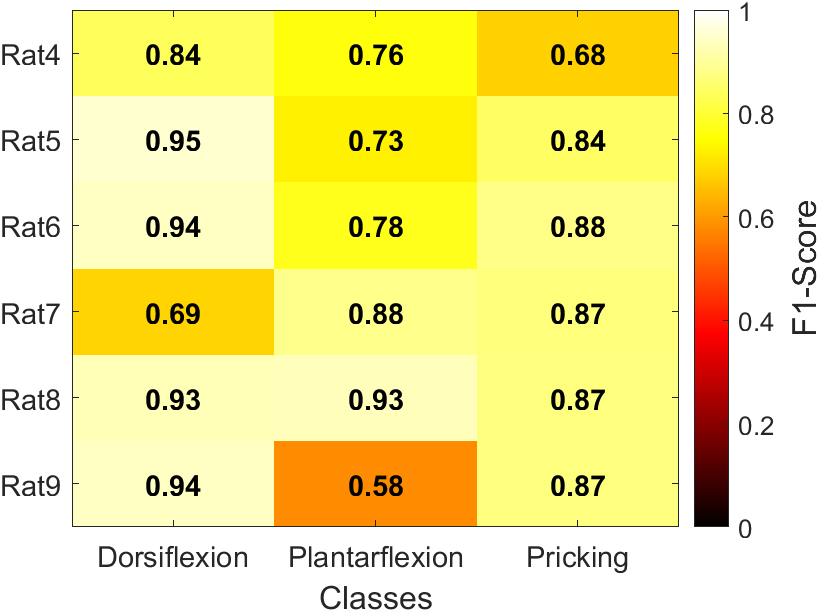}  
        \subcaption{}  
        \label{fig:subplot2}
    \end{minipage}
    \begin{minipage}{0.65\columnwidth}
        \centering
        \includegraphics[width=0.8\columnwidth]{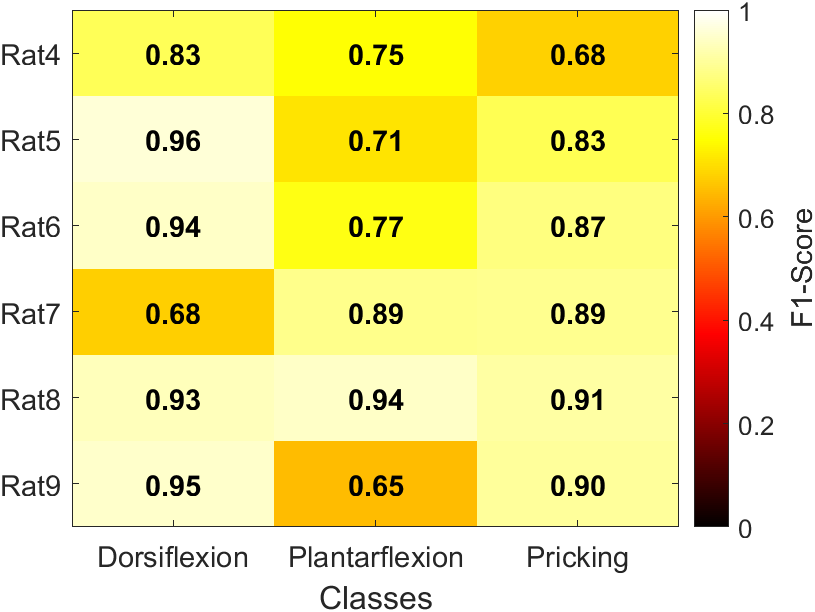}  
        \subcaption{}  
        \label{fig:subplot3}
    \end{minipage}
    \caption{Classification F1-scores across different classes and subjects using (a) ESCAPE and (b) MobilESCAPE-Net.}
    \label{fig:each_rat_n_class_f1}
\end{figure}

\subsection{MobilESCAPE-NET vs. ESCAPE-NET}
The comparison is conducted in terms of classification F1-score, accuracy, and complexity, considering all the rats detailed in Table \ref{tab:snr_matrix}.



Figure~\ref{fig:each_rat_n_class_f1} demonstrates the robustness of MobileESCAPE-Net, in terms of F1-score, to variations across rats (and SNR, as detailed in Table~\ref{tab:snr_matrix}) and different stimuli. The model shows standard deviations of 0.10, 0.10, and 0.08 for dorsiflexion, plantarflexion, and pricking, respectively. Notably, the standard deviation for the plantarflexion stimulus across different animals decreases from 
0.12 to 0.10 when using MobileESCAPE-Net. Additionally, both accuracy and F1-score remain comparable between the two models, except in cases of low SNR (e.g., Rat 9 with Plantarflexion, with $\mathrm{SNR} = 0.41 \mathrm{dB}$), where MobilESCAPE-Net outperforms ESCAPE-NET of $10.7\%$.
\begin{figure}[h]
        \includegraphics[width=0.58\columnwidth]{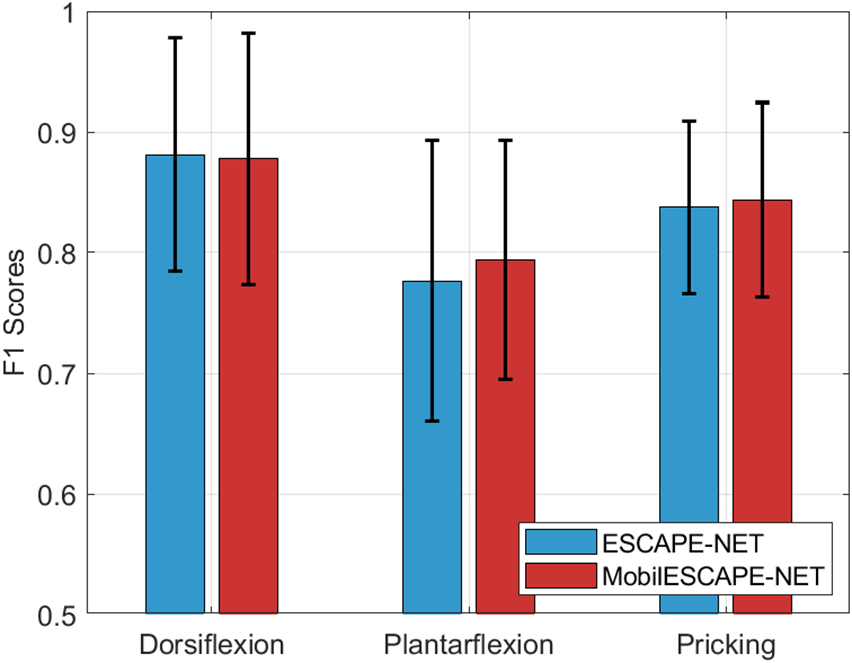}  
        \centering\caption{Comparison of F1-Scores for dorsiflexion, plantarflexion, and pricking classification using ESCAPE-NET and MobilESCAPE-NET. Error bars represent the standard deviation.}  
        \label{fig:f1score_each_class}
\end{figure}

Figure~\ref{fig:f1score_each_class} compares class-wise F1-scores across all rats, highlighting plantarflexion as the most challenging class, with ESCAPE-Net and MobileESCAPE-Net scoring 0.776 and 0.794, respectively. MobileESCAPE-Net shows slight improvements in plantarflexion (+1.8\%) and pricking (+0.6\%) but a minor drop (-0.4\%) in dorsiflexion. Table~\ref{tab:acc_f1_total} further validates its performance, showing comparable accuracy and macro F1-score to ESCAPE-Net, with results averaged across animals and classes.

While both models perform comparably, the key advantage of MobileESCAPE-Net is its ability to achieve this performance with just 67,843 trainable parameters and 82.8M FLOPs, representing a 99.92\% reduction in parameters and a 92.47\% reduction in FLOPs compared to ESCAPE-Net. These values, shown in Table~\ref{tab:model_complexity}, provide a clear comparison of model complexity and computational efficiency. This significant reduction is particularly critical for real-time applications in implantable devices for peripheral nerve function restoration, where computational resources are limited. MobileESCAPE-Net's efficiency enables deployment in such resource-constrained systems, allowing fast and reliable processing without sacrificing accuracy, making it well-suited for real-time ENG stimuli classification in implantable settings.

\begin{table}[t]
\centering
\caption{Average Classification Accuracies and Corresponding F1-Scores.}
\label{tab:ref_montages}
\resizebox{0.95 \columnwidth}{!}{ 
\begin{tabular}{lcc} 
\toprule
\textbf{Model} & \textbf{Classification Accuracy (\%)} & \textbf{Macro F1-score} \\
\midrule

ESCAPE-NET & 85.71 $\pm$ 5.34 & 0.83 $\pm$ 0.11 \\
MobilESCAPE-NET & 85.82 $\pm$ 5.07 & 0.84 $\pm$ 0.10 \\

\bottomrule
\end{tabular}
}
\label{tab:acc_f1_total}
\end{table}
\vspace{-0.5em}

\begin{table}[h!]
\centering
\caption{Model Complexity.}
\label{tab:ref_montages}
\resizebox{0.7\columnwidth}{!}{ 
\begin{tabular}{lcc} 
\toprule
\textbf{Model} & \textbf{Parameters}& \textbf{FLOPs} \\
\midrule
ESCAPE-NET & 91837635 & 1.1G \\
MobilESCAPE-NET & 67843 & 82.8M \\
\bottomrule
\end{tabular}
}
\label{tab:model_complexity}
\end{table}

\section{Conclusions}\label{sect:conclusion}

Accurate and real-time classification of ENG signals remains a key challenge in addressing peripheral nerve injuries, particularly in the context of implantable devices where computational resources are limited. This work presents MobilESCAPE-Net, an optimized deep learning architecture specifically designed to meet the low-latency and low-complexity requirements of such systems. MobilESCAPE-Net achieves comparable accuracy and F1-score to the state-of-the-art ESCAPE-Net while significantly reducing trainable parameters by 99.9\% and floating-point operations (FLOPs) by 92.47\%. These reductions enable faster inference and make the model suitable for real-time deployment. The proposed architecture demonstrates strong potential for integration into ND\&S systems.

\section*{Acknowledgment}
The analysis in this paper relies on the dataset which has been created by Zariffa et al. (2023) and it is available at https://doi.org/10.5683/SP3/JRZDDR



\bibliographystyle{IEEEtran}
\bibliography{bibliography}

\begin{thebibliography}{10}
\providecommand{\url}[1]{#1}
\csname url@samestyle\endcsname
\providecommand{\newblock}{\relax}
\providecommand{\bibinfo}[2]{#2}
\providecommand{\BIBentrySTDinterwordspacing}{\spaceskip=0pt\relax}
\providecommand{\BIBentryALTinterwordstretchfactor}{4}
\providecommand{\BIBentryALTinterwordspacing}{\spaceskip=\fontdimen2\font plus
\BIBentryALTinterwordstretchfactor\fontdimen3\font minus \fontdimen4\font\relax}
\providecommand{\BIBforeignlanguage}[2]{{%
\expandafter\ifx\csname l@#1\endcsname\relax
\typeout{** WARNING: IEEEtran.bst: No hyphenation pattern has been}%
\typeout{** loaded for the language `#1'. Using the pattern for}%
\typeout{** the default language instead.}%
\else
\language=\csname l@#1\endcsname
\fi
#2}}
\providecommand{\BIBdecl}{\relax}
\BIBdecl

\bibitem{[4]caillaud2019peripheral}
M.~Caillaud, L.~Richard, J.-M. Vallat, A.~Desmouli{\`e}re, and F.~Billet, ``Peripheral nerve regeneration and intraneural revascularization,'' \emph{Neural regeneration research}, vol.~14, no.~1, p.~24, 2019.

\bibitem{paperasp}
A.~Coviello, C.~Cavigliano, V.~Tasso, E.~T. Tavassi, Y.~Giacalone, and M.~Magarini, ``Emerging peripheral nerve injuries recovery: advanced nerve-cuff electrode model interface for implantable devices,'' in \emph{2024 Global Conference on Wireless and Optical Technologies (GCWOT)}.\hskip 1em plus 0.5em minus 0.4em\relax IEEE, 2024, pp. 1--7.

\bibitem{[5]Denison65}
\BIBentryALTinterwordspacing
T.~Denison and M.~J. Morrell, ``Neuromodulation in 2035,'' \emph{Neurology}, vol.~98, no.~2, pp. 65--72, 2022. [Online]. Available: \url{https://n.neurology.org/content/98/2/65}
\BIBentrySTDinterwordspacing

\bibitem{[10][11]Controller}
T.~R. Farrell and R.~F. Weir, ``The optimal controller delay for myoelectric prostheses,'' \emph{IEEE Transactions on Neural Systems and Rehabilitation Engineering}, vol.~15, no.~1, pp. 111--118, 2007.

\bibitem{martinek2021advanced}
R.~Martinek, M.~Ladrova, M.~Sidikova, R.~Jaros, K.~Behbehani, R.~Kahankova, and A.~Kawala-Sterniuk, ``Advanced bioelectrical signal processing methods: Past, present, and future approach—part iii: Other biosignals,'' \emph{Sensors}, vol.~21, no.~18, p. 6064, 2021.

\bibitem{hosseini2020review}
M.-P. Hosseini, A.~Hosseini, and K.~Ahi, ``A review on machine learning for eeg signal processing in bioengineering,'' \emph{IEEE reviews in biomedical engineering}, vol.~14, pp. 204--218, 2020.

\bibitem{10530475}
A.~Coviello, F.~Linsalata, U.~Spagnolini, and M.~Magarini, ``Artificial neural networks-based real-time classification of eng signals for implanted nerve interfaces,'' \emph{IEEE Journal on Selected Areas in Communications}, vol.~42, no.~8, pp. 2080--2095, 2024.

\bibitem{kohSelectivePeripheralNerve2020journalArticle}
R.~G.~L. Koh, M.~Balas, A.~I. Nachman, and J.~Zariffa, ``Selective peripheral nerve recordings from nerve cuff electrodes using convolutional neural networks,'' \emph{Journal of Neural Engineering}, vol.~17, no.~1, p. 016042, Jan. 2020.

\bibitem{lawhern2018eegnet}
V.~J. Lawhern, A.~J. Solon, N.~R. Waytowich, S.~M. Gordon, C.~P. Hung, and B.~J. Lance, ``Eegnet: a compact convolutional neural network for eeg-based brain--computer interfaces,'' \emph{Journal of neural engineering}, vol.~15, no.~5, p. 056013, 2018.

\bibitem{sanei2021eeg}
S.~Sanei and J.~A. Chambers, \emph{EEG signal processing and machine learning}.\hskip 1em plus 0.5em minus 0.4em\relax John Wiley \& Sons, 2021.

\bibitem{SP3/JRZDDR_2023}
\BIBentryALTinterwordspacing
J.~Zariffa, ``{Replication Data for: Koh, Ryan GL, Adrian I. Nachman, and José Zariffa. ``Classification of naturally evoked compound action potentials in peripheral nerve spatiotemporal recordings.'' Scientific reports 9:11145, 2019.}'' 2023. [Online]. Available: \url{https://doi.org/10.5683/SP3/JRZDDR}
\BIBentrySTDinterwordspacing

\bibitem{struijkTripolarNerveCuff1995conferencePaper}
J.~Struijk and M.~Thomsen, ``Tripolar nerve cuff recording: Stimulus artifact, {{EMG}} and the recorded nerve signal,'' in \emph{Proceedings of 17th {{International Conference}} of the {{Engineering}} in {{Medicine}} and {{Biology Society}}}, vol.~2, Sep. 1995, pp. 1105--1106 vol.2.

\bibitem{howardMobileNetsEfficientConvolutional2017preprinta}
A.~G. Howard, M.~Zhu, B.~Chen, D.~Kalenichenko, W.~Wang, T.~Weyand, M.~Andreetto, and H.~Adam, ``{{MobileNets}}: {{Efficient Convolutional Neural Networks}} for {{Mobile Vision Applications}},'' Apr. 2017.

\bibitem{linNetworkNetwork2014preprint}
M.~Lin, Q.~Chen, and S.~Yan, ``Network {{In Network}},'' Mar. 2014.

\end{thebibliography}

\end{document}